\documentclass{webofc}
\usepackage{amsmath,graphicx,multirow,tabularx,physics}
\usepackage[varg]{txfonts}   
\usepackage{hyperref}
\usepackage{url}
\usepackage{wrapfig}
\usepackage{subcaption} 
\hypersetup{colorlinks=true,citecolor=blue,urlcolor=blue,linkcolor=blue}
\begin{document}
\title{Higher order net-baryon number cumulants and baryon-strangeness correlations: Comparing QCD results on the pseudo-critical line with RHIC-BES II results on the freeze-out line}
\author{\firstname{Jishnu} \lastname{Goswami}\inst{1}\fnsep\thanks{\email{jishnu@physik.uni-bielefeld.de}} \and
        \firstname{Frithjof} \lastname{Karsch}\inst{1}\fnsep\thanks{\email{karsch@physik.uni-bielefeld.de}}
}

\institute{Fakult\"at f\"ur Physik, Universit\"at Bielefeld, Bielefeld, 33615, Germany}
\abstract{
We present lattice QCD results for ratios of net-baryon number cumulants along the 
pseudo-critical line and compare them with STAR measurements from the RHIC BES-II 
program. The ratio of first and second order cumulants, $R_{12}^B$, agrees well with 
corresponding net-proton number cumulants down to 
$\sqrt{s_{NN}}=11.5$~GeV or baryon chemical potentials $\mu_B/T\le 2$. 
Likewise higher-order cumulant ratios, $R_{31}^B$ and $R_{42}^B$, 
show no sign for the existence of a critical point in the parameter range explored with these
cumulant ratios. A QCD critical point is unlikely to occur within the BES-II range in collider mode.
Moreover, the results demonstrate that a non-interacting HRG description breaks down 
for $\mu_B/T > 1$.  We further analyze baryon–strangeness correlations normalized by strangeness fluctuations, finding consistency with STAR data at large beam energies but deviations at lower energies. Comparisons of electric-charge and strangeness correlations with STAR and ALICE data also show agreement at high energies, while the deviations at lower energies emphasize the role of unobserved strange resonances and the need for controlled feed-down corrections in baryon–strangeness correlations.  
}
\maketitle
\section{Introduction}
\label{intro}
In this proceeding contribution we present updated results on net-baryon number cumulants 
and on correlations of baryon and strangeness number fluctuations, normalized by 
second-order strangeness fluctuations, along the pseudo-critical line of 
(2+1)-flavor QCD. The calculations are based on high-statistics data generated 
by the HotQCD collaboration, extending up to eighth order in the Taylor expansion 
of the pressure. We compare our results with point-like, non-interacting hadron 
resonance gas (HRG) model calculations as well as with Beam Energy Scan (BES-II) 
data obtained by the STAR collaboration at RHIC in the United States.  

In our comparisons to HRG model calculations, we use two resonance lists:  
PDG-HRG, which includes only the 3- and 4-star states listed in the PDG booklet, and  
QMHRG2020~\cite{Bollweg:2021vqf}, which additionally incorporates 1- and 2-star resonances from the PDG booklet, 
 together with further resonances predicted by quark model calculations.  
In the latter case, particular care has been taken to avoid double counting of states.  

Throughout this proceeding, $T_{pc}(\mu_B)$ denotes the pseudo-critical line as a function of baryon chemical potential $\mu_B$ obtained from determinations of the maximum of the chiral susceptibility; 
$T_{pc}^0$ refers to the pseudo-critical temperature at vanishing chemical potentials. 
\vspace{-0.2cm}
\section{QCD critical point constraints from the pseudo-critical line  of (2+1)-flavor QCD}
\label{sec-1}
The pressure of (2+1)-flavor QCD at finite baryon, electric charge, and strangeness chemical 
potentials can be 
expressed in a Taylor expansion around $\vec{\mu}=(\mu_B,\mu_Q,\mu_S)=\vec{0}$,  
\begin{equation}
\frac{P}{T^4} = \sum_{i+j+k=\rm even} 
  \frac{\chi_{ijk}^{BQS}}{i!\,j!\,k!}\,
  \hat\mu_B^i \hat\mu_Q^j \hat\mu_S^k,
\quad
\chi_{ijk}^{BQS} = \frac{1}{VT^3}\,
\frac{\partial^{\,i+j+k} \ln Z}
     {\partial\hat\mu_B^i \partial\hat\mu_Q^j \partial\hat\mu_S^k}\Big|_{\mu=0}\; ,
\label{chiTaylor}
\end{equation}
with $\hat\mu_X\equiv \mu_X/T$, for $X=B,\ Q,\ S$.
Here the generalized susceptibilities $\chi_{ijk}^{BQS}$ encode fluctuations and 
correlations of conserved charges and are the basic observables that connect lattice 
QCD calculations with heavy-ion experiments.  

In heavy-ion collisions, constraints on the net charge densities $n_Q/n_B = 0.4$ and 
vanishing strangeness, $n_S=0$, fix the values of $\mu_Q$ and $\mu_S$ as functions of $\mu_B$
\cite{Bazavov:2020bjn}.
The expansion of cumulants and other thermodynamic observables thus effectively reduces to a 
one-parameter series in $\hat\mu_B$. The pseudo-critical line can be parameterized as  
\begin{equation}
  T_{pc}(\mu_B) = T_{pc}^0\left[1 - \kappa_2 \hat\mu_B^{\,2} + \kappa_4 \hat\mu_B^{\,4}+{\cal O}(\hat\mu_B^6)\right],
  \label{Tpc}
\end{equation}
with $T_{pc}^0 = (156.5 \pm 1.5)$~MeV, $\kappa_2 = 0.012(4)$, and $\kappa_4 = 0.000(4)$ 
for $n_S=0$~\cite{HotQCD:2018pds}. The coefficient $\kappa_2$ has been shown to be only
weakly quark mass dependent, {\it i.e.} $\kappa_2 =0.015(1)$ for vanishing light quark masses \cite{Ding:2024sux}, while the chiral phase transition temperature drops by about 25 MeV.
It has been found to be $T_c^0= 132^{+3}_{-6}$ MeV \cite{HotQCD:2019xnw}.
If present at all, the QCD critical point (cep) is expected to show up at 
temperature  $T^{cep}<T_c^0$
\cite{Halasz:1998qr,Hatta:2002sj,Ding:2024sux}. The chiral phase transition temperature thus  
puts a bound on $T^{cep}$ for the
location of a possible CEP and as such also gives a lower bound on $\mu_B^{cep}$. Assuming that the 
curvature of $T_{pc}(\mu_B)$ is well described by Eq.~\ref{Tpc} up to $\hat\mu_B\simeq 3.6$, where
$T_{pc}(\mu_B)$ would reach a temperature of about 132 MeV, one finds
\begin{equation}
  T_{c}^{cep}(\mu_B) < 132^{+3}_{-6}~\mathrm{MeV}, \qquad \mu_B^{cep} > 477^{+30}_{-20}~\mathrm{MeV}.
\end{equation}
An even more stringent bound is obtained when assuming that a possibly existing tri-critical point
in the chiral limit does not occur for $\hat\mu_B<2$. In that case one concludes that 
no critical point can appear at physical values of the quark masses for
$T > 125$~MeV and $\mu_B< 510$~MeV. 
This suggests that if a 
critical point exists at all, it must lie at values of $\mu_B$ beyond those 
currently reached by BES-II in the collider mode. In the following sections we will show that fluctuation 
observables, especially skewness and kurtosis, are consistent with this conclusion.
\vspace{-0.8cm}
\section{Net-baryon number fluctuations on the pseudo-critical line}
\begin{figure}[t]
\centering
\includegraphics[width=5cm,clip]{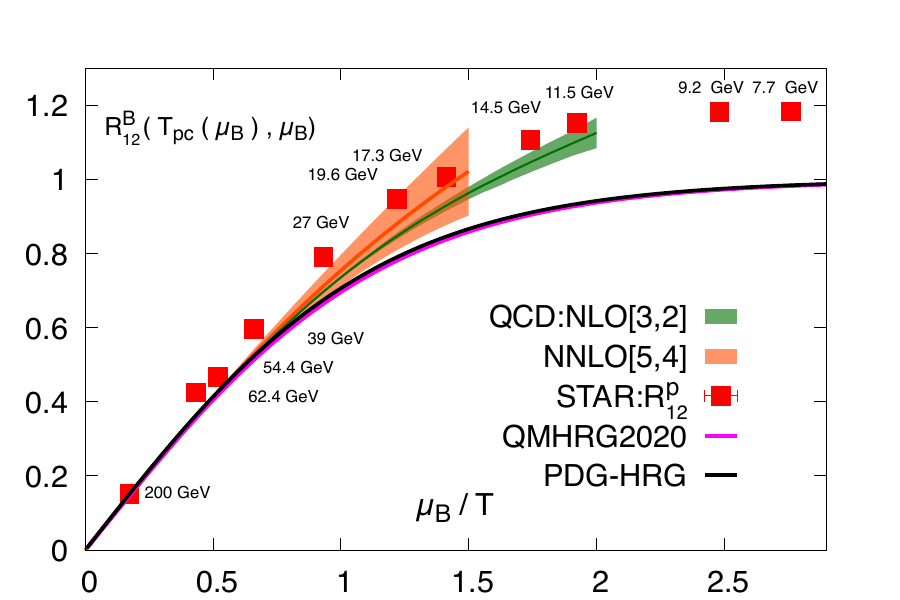}
\includegraphics[width=6.6cm,clip]{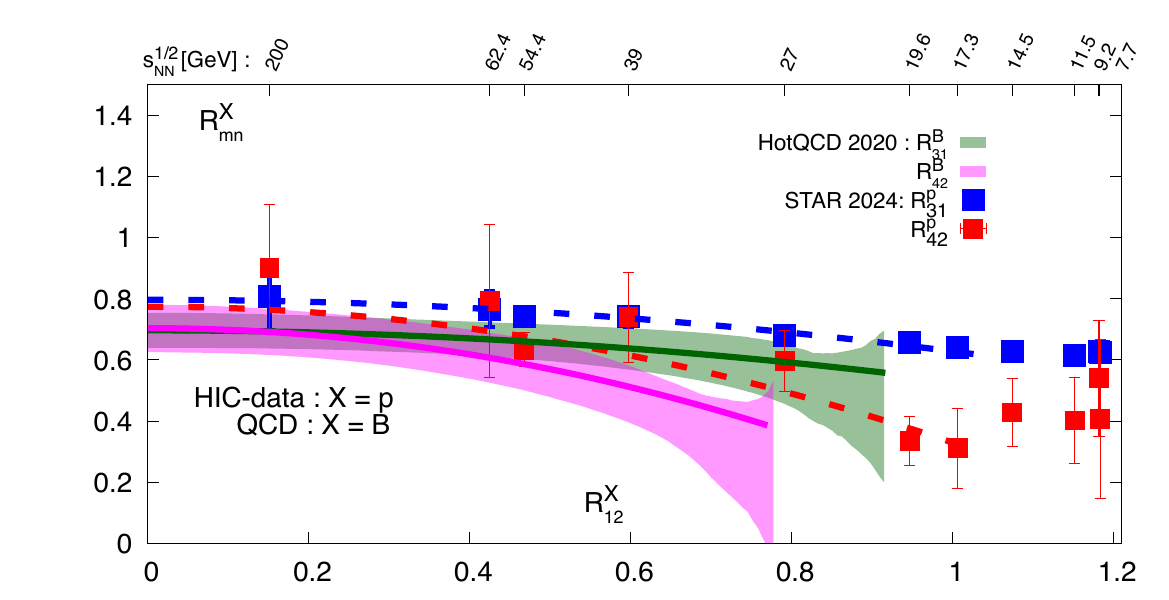}
\caption{{\it Left} The ratio $R_{12}^B$ obtained in QCD and HRG-model calculations 
as function $\mu_B/T$ on the $T_{pc}$. {\it Right:} Skewness ($R_{31}^B$) and
kurtosis ($R_{42}^B$) ratios
as function of $R_{12}^B$. 
Also shown are  the corresponding STAR results for net-proton number fluctuations.}
\label{fig:baryon_number}  
\vspace{-0.9cm}
\end{figure}

In Fig.~\ref{fig:baryon_number}, we present the ratio of mean and variance of the net-baryon number distributions, 
$R_{12}^B = \chi_1^B / \chi_2^B$, in (2+1)-flavor QCD as a function of $\mu_B/T$ along the pseudo-critical line $T_{pc}(\mu_B)$. 
We find that $R_{12}^B$ and the corresponding $R_{12}^p = \chi_1^p / \chi_2^p$ of net-proton number fluctuations, taken from BES-II STAR data \cite{STAR:2025zdq}, 
are consistent with each other down to beam energies $\sqrt{s_{NN}} \simeq 11.5$~GeV. 
At larger baryon chemical potentials, both $R_{12}^B$ and $R_{12}^p$ exceed unity, 
in clear disagreement with point-like, non-interacting HRG-model predictions, 
which gives $R_{12}^B \leq 1$ and $R_{12}^p \leq 1$ for any value of $\mu_B$. This indicates the presence of more complicated baryonic interactions that are absent in a point-like, non-interacting HRG.

For higher-order cumulants, we find that QCD results for skewness ratio $R_{31}^B\equiv \chi_3^B/\chi_1^B$, and kurtosis ratio, $R_{42}^B\equiv \chi_4^B/\chi_2^B$, are consistent with 
STAR data down to $\sqrt{s_{NN}} \simeq 19.6$~GeV and $\sqrt{s_{NN}} \simeq 27$~GeV, respectively. 
In contrast, these ratios are unity in calculations with a point-like, non-interacting HRG-model. 
While these ratios approach unity also in QCD at low temperatures, 
they decrease by about $20\text{--}30\%$ below unity already at $T_{pc}^0$. 
The point-like, non-interacting HRG-model thus fails to describe these observables even $\mu_B/T=0$.
The agreement of $R_{12}^B$ with STAR net-proton results down to $\sqrt{s_{NN}}\simeq 11.5$~GeV, 
together with the absence of any non-monotonicity in skewness and kurtosis ratios, 
provides further evidence against a the existence of CEP in the ($T,\mu_B)$ range probed in heavy ion
collisions with beam energies $\sqrt{s_{NN}}\ge 11.5$~GeV.
\vspace{-0.2cm}
\section{Baryon--strangeness number correlations on the pseudo-critical line}
\begin{figure*}
\centering
\vspace*{1cm}
\includegraphics[width=5cm,clip]{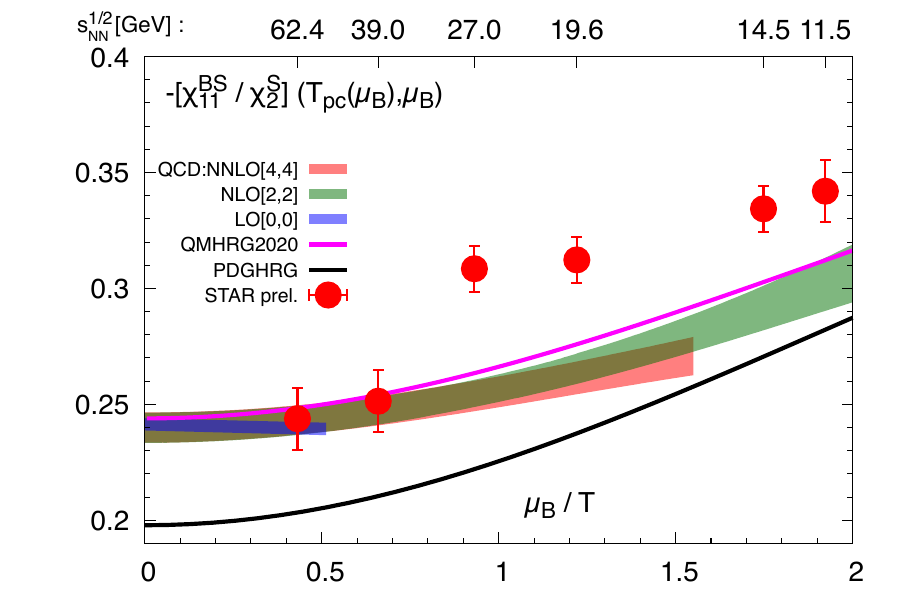}
\includegraphics[width=5cm,clip]{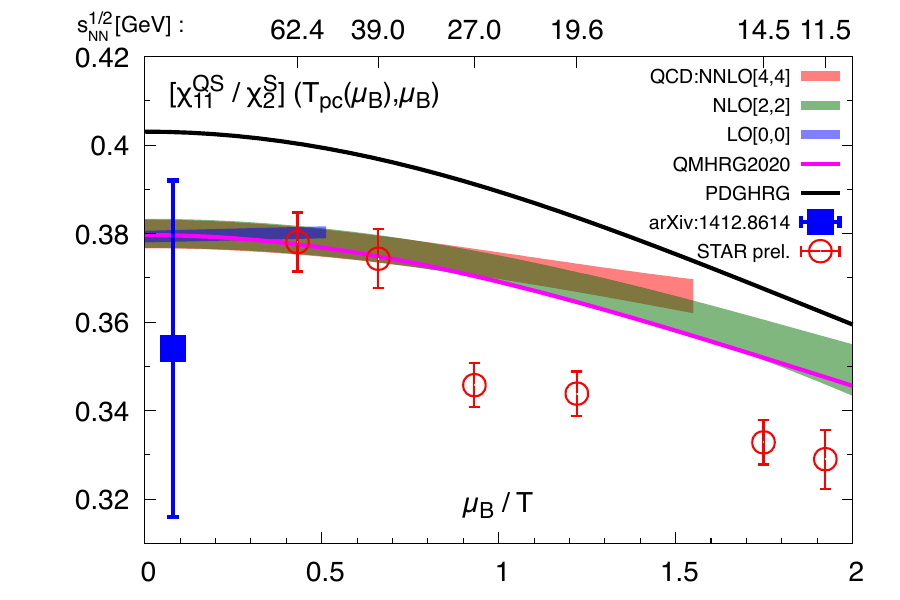}
\caption{Baryon–strangeness (left) and electric charge–strangeness (right) correlations normalized 
by $\chi_2^S$ along $T_{pc}(\mu_B)$~\cite{Bollweg:2024epj}. Lattice QCD results are compared with preliminary heavy ion collision
data from STAR presented at CPOD 2024 \cite{STAR-CPOD}, and HRG model calculations using the QMHRG2020 list of hadrons. Also shown is a data point based on an analysis of data obtained by the ALICE collaboration \cite{Braun-Munzinger:2014lba}.}
\label{fig:baryon_strangeness}
\vspace{-0.9cm}
\end{figure*}

In Fig.~\ref{fig:baryon_strangeness} (left), we show the cumulant ratio $\chi_{11}^{BS}/\chi_{2}^S$, describing correlations of net-baryon and net-strangeness numbers, $\chi_{11}^{BS}$, normalized by the 
second-order strangeness fluctuations $\chi_{2}^S$. Results are shown along the pseudo-critical line $T_{pc}(\mu_B)$. In our previous publications, we studied truncation effects on Taylor series for second-order conserved-charge cumulants and concluded that our results are reliable for $\mu_B/T \leq 1.5$~\cite{Bollweg:2022rps,Bollweg:2024epj}.
In this range of $\mu_B/T$, QCD results agree well with STAR data only for $\sqrt{s_{NN}} \geq 39$~GeV, while significant deviations appear for $\sqrt{s_{NN}} \leq 27$~GeV. We also find consistency between QCD and QMHRG2020 up to $\mu_B/T \simeq 2$, while sizeable deviations from PDG-HRG results exist for all $\mu_B/T\ge 0$. Since QMHRG2020 includes many unobserved hadrons, this may point to the role of additional strange hadrons when constructing this observable from experimental data.
In Fig.~\ref{fig:baryon_strangeness} (right), we present corresponding results for the ratio of electric charge--strangeness 
correlations, $\chi_{11}^{QS}$, and $\chi_{2}^S$,
\begin{equation}
2 \frac{\chi_{11}^{QS}(T,\vec{\mu})}{\chi_{2}^S(T,\vec{\mu})} - 
\frac{\chi_{11}^{BS}(T,\vec{\mu})}{\chi_{2}^S(T,\vec{\mu})}= 1 + 
\frac{\Delta^{BQS}(T,\vec{\mu})}{\chi_{2}^S(T,\vec{\mu})}\, ,
\label{BS-QS-rel}
\end{equation}
with $\Delta^{BQS}=0$ for strangeness neutral, isospin symmetric matter ($n_S=0, n_Q/n_B=0.5$).
For conditions relevant to heavy-ion collisions 
($n_S=0$, $n_Q/n_B = 0.4$), the deviation from unity ($\Delta^{BQS}>0$) is less than $0.5\%$. 
Accordingly, the comparison of $\chi_{11}^{QS}/\chi_{2}^S$ with STAR and QCD data shows the same trends observed for 
$\chi_{11}^{BS}/\chi_{2}^S$. 
 Fig.~\ref{fig:baryon_strangeness} also includes the ratio $\chi_{11}^{QS}/\chi_{2}^S$ obtained in \cite{Braun-Munzinger:2014lba} from strange 
particle yields measured by the ALICE Collaboration at the LHC~\cite{ALICE:2013mez,ALICE:2014jbq}. The second-order cumulants 
$\chi_{2}^S$, $\chi_{11}^{BS}$, and $\chi_{11}^{QS}$ were constructed from the measured yields, with feed-down 
corrections from $\phi$-mesons and neutral kaons~\cite{Braun-Munzinger:2014lba}. 
These cumulants satisfy 
Eq.~\ref{BS-QS-rel} 
up to uncertainties arising from feed-down contributions. As emphasized in \cite{Braun-Munzinger:2014lba}, some decay channels contributing to $\chi_{11}^{BS}$ are not known experimentally, 
and feed-down from higher kaon resonances as well as additional strange hadrons is difficult to control. 
Thus, a precise experimental determination of $\chi_{11}^{BS}$ will remain challenging until these resonance 
contributions and their decay channels are better constrained.  
\vspace{-0.2cm}
\section{Summary}
\vspace{-0.2cm}
We presented current constraints on the location of a possibly existing critical endpoint lattice QCD 
at non-vanishing chemical potential arising from studies of the QCD chiral phase transition temperature
and discussed updated results for net-baryon number cumulants and baryon–strangeness 
correlations along the pseudo-critical line of (2+1)-flavor QCD.

\vspace{0.2cm}
\noindent
{\it \bf{Acknowledgements}}--This work was supported by the Deutsche Forschungsgemeinschaft
(DFG, German Research Foundation) Proj. No. 315477589-TRR 211
and the consortium PUNCH4NFDI
supported by the Deutsche Forschungsgemeinschaft (DFG, German Research Foundation) with project number 460248186 (PUNCH4NFDI).
\vspace{-0.5cm}
\bibliography{bibliography.bib}

\begin{thebibliography}{14}

\bibitem{Bollweg:2021vqf}
D.~Bollweg et~al. (HotQCD), Phys. Rev. D \textbf{104} (2021),
  \texttt{2107.10011}.

\bibitem{Bazavov:2020bjn}
A.~Bazavov et~al., Phys. Rev. D \textbf{101}, 074502 (2020),
  \texttt{2001.08530}.

\bibitem{HotQCD:2018pds}
A.~Bazavov et~al. (HotQCD), Phys. Lett. B \textbf{795}, 15 (2019),
  \texttt{1812.08235}.

\bibitem{Ding:2024sux}
H.T. Ding et~al., Phys. Rev. D \textbf{109}, 114516 (2024),
  \texttt{2403.09390}.

\bibitem{HotQCD:2019xnw}
H.T. Ding et~al. (HotQCD), Phys. Rev. Lett. \textbf{123}, 062002 (2019).

\bibitem{Halasz:1998qr}
A.M. Halasz,  et~al., Phys. Rev. D \textbf{58}, 096007 (1998),
  \texttt{hep-ph/9804290}.

\bibitem{Hatta:2002sj}
Y.~Hatta, T.~Ikeda, Phys. Rev. D \textbf{67}, 014028 (2003),
  \texttt{hep-ph/0210284}.

\bibitem{STAR:2025zdq}
B.E. Aboona et~al. (STAR), Phys. Rev. Lett. \textbf{135}, 142301 (2025),
  \texttt{2504.00817}.

\bibitem{Bollweg:2022rps}
D.~Bollweg et~al. (HotQCD), Phys. Rev. D \textbf{105}, 074511 (2022),
  \texttt{2202.09184}.

\bibitem{Bollweg:2024epj}
D.~Bollweg et~al., Phys. Rev. D \textbf{110}, 054519 (2024),
  \texttt{2407.09335}.

\bibitem{STAR-CPOD}
H.~Feng (for the STAR Collaboration), {Baryon-Strangeness Correlations in Au+Au
  Collisions at RHIC-STAR, 15th Workshop on Critical Point and Onset of
  Deconfinement, Berkeley, May 20-24, 2024.}

\bibitem{Braun-Munzinger:2014lba}
P.~Braun-Munzinger, A.~Kalweit, K.~Redlich, J.~Stachel, Phys. Lett. B
  \textbf{747}, 292 (2015).

\bibitem{ALICE:2013mez}
B.~Abelev et~al. (ALICE), Phys. Rev. C \textbf{88}, 044910 (2013),
  \texttt{1303.0737}.

\bibitem{ALICE:2014jbq}
B.B. Abelev et~al. (ALICE), Phys. Rev. C \textbf{91}, 024609 (2015),
  \texttt{1404.0495}.

\end{thebibliography}
\end{document}